\documentclass[reprint,aps,amsmath,amssymb,longbibliography,floatfix,superscriptaddress,10pt,tightenlines]{revtex4-1}

\usepackage[english]{babel}
\usepackage{bm,graphicx,braket,color,booktabs,fixmath}
\usepackage[hidelinks,colorlinks=true,citecolor=black,linkcolor=black,urlcolor=black]{hyperref}

\begin{document}

\title{Reprogrammable Electro-Optic Nonlinear Activation Functions \\ for Optical Neural Networks}

\author{Ian A. D. Williamson}
\email{iwill@stanford.edu}
\affiliation{Department of Electrical Engineering and Ginzton Laboratory, Stanford University, Stanford, CA 94305, USA}
\author{Tyler W. Hughes}
\affiliation{Department of Applied Physics and Ginzton Laboratory, Stanford University, Stanford, CA 94305, USA}
\author{Momchil Minkov}
\affiliation{Department of Electrical Engineering and Ginzton Laboratory, Stanford University, Stanford, CA 94305, USA}
\author{Ben Bartlett}
\affiliation{Department of Applied Physics and Ginzton Laboratory, Stanford University, Stanford, CA 94305, USA}
\author{Sunil Pai}
\affiliation{Department of Electrical Engineering and Ginzton Laboratory, Stanford University, Stanford, CA 94305, USA}
\author{Shanhui~Fan}
\email{shanhui@stanford.edu}
\affiliation{Department of Electrical Engineering and Ginzton Laboratory, Stanford University, Stanford, CA 94305, USA}

\begin{abstract}
    We introduce an electro-optic hardware platform for nonlinear activation functions in optical neural networks. The optical-to-optical nonlinearity operates by converting a small portion of the input optical signal into an analog electric signal, which is used to intensity-modulate the original optical signal with no reduction in processing speed. Our scheme allows for complete nonlinear on-off contrast in transmission at relatively low optical power thresholds and eliminates the requirement of having additional optical sources between each layer of the network. Moreover, the activation function is reconfigurable via electrical bias, allowing it to be programmed or trained to synthesize a variety of nonlinear responses. Using numerical simulations, we demonstrate that this activation function significantly improves the expressiveness of optical neural networks, allowing them to perform well on two benchmark machine learning tasks: learning a multi-input exclusive-OR (XOR) logic function and classification of images of handwritten numbers from the MNIST dataset. The addition of the nonlinear activation function improves test accuracy on the MNIST task from 85\% to 94\%.
\end{abstract}

\maketitle

\section{Introduction}

In recent years, there has been significant interest in alternative computing platforms specialized for high performance and efficiency on machine learning tasks.
For example, graphical processing units (GPUs) have demonstrated peak performance with trillions of floating point operations per second (TFLOPS) when performing matrix multiplication, which is several orders of magnitude larger than general-purpose digital processors such as CPUs \cite{pallipuram_comparative_2012}.
Moreover, analog computing has been explored for achieving high performance because it is not limited by the bottlenecks of sequential instruction execution and memory access \cite{shainline_superconducting_2017, shastri_principles_2018, coarer_alloptical_2018, chang_hybrid_2018, colburn_optical_2019}.

Optical hardware platforms are particularly appealing for computing and signal processing due to their ultra-large signal bandwidths, low latencies, and reconfigurability \cite{capmany_microwave_2007, marpaung_integrated_2013, ghelfi_fully_2014}. 
They have also gathered significant interest in machine learning applications, such as artificial neural networks (ANNs). 
Nearly three decades ago, the first optical neural networks (ONNs) were proposed based on free-space optical lens and holography setups \cite{abu-mostafa_optical_1987, psaltis_holography_1990}. 
More recently, ONNs have been implemented in chip-integrated photonic platforms \cite{shen_deep_2017} using programmable waveguide interferometer meshes which perform matrix-vector multiplications \cite{miller_selfconfiguring_2013}. 
In theory, the performance of such systems is competitive with digital computing platforms because they may perform matrix-vector multiplications in constant time with respect to the matrix dimension. 
In contrast, matrix-vector multiplication has a quadratic time complexity on a digital processor. 
Other approaches to performing matrix-vector multiplications in chip-integrated ONNs, such as microring weight banks and photodiodes, have also been proposed \cite{tait_neuromorphic_2017}.

Nonlinear activation functions play a key role in ANNs by enabling them to learn complex mappings between their inputs and outputs. 
Whereas digital processors have the expressiveness to trivially apply nonlinearities such as the widely-used \texttt{sigmoid}, \texttt{ReLU}, and \texttt{tanh} functions, the realization of nonlinearities in optical hardware platforms is more challenging. 
One reason for this is that optical nonlinearities are relatively weak, necessitating a combination of large interaction lengths and high signal powers, which impose lower bounds on the physical footprint and the energy consumption, respectively. 
Although it is possible to resonantly enhance optical nonlinearities, this comes with an unavoidable trade-off in reducing the operating bandwidth, thereby limiting the information processing capacity of an ONN. 
Additionally, maintaining uniform resonant responses across many elements of an optical circuit necessitates additional control circuitry for calibrating each element \cite{radulaski_thermally_2018}.

A more fundamental limitation of optical nonlinearities is that their responses tend to be fixed during device fabrication. 
This limited tunability of the nonlinear optical response prevents an ONN from being reprogrammed to realize different forms of nonlinear activation functions, which may be important for tailoring ONNs for different machine learning tasks. 
Similarly, a fixed nonlinear response may also limit the performance of very deep ONNs with many layers of activation functions since the optical signal power drops below the activation threshold, where nonlinearity is strongest, in later layers due to loss in previous layers. 
For example, with optical saturable absorption from 2D materials in waveguides, the activation threshold is on the order of 1-10 mW \cite{bao_monolayer_2011, park_monolayer_2015, jiang_low-dimensional_2018}, meaning that the strength of the nonlinearity in each subsequent layer will be successively weaker as the transmitted power falls below the threshold.

In light of these challenges, the ONN demonstrated in Ref. \citenum{shen_deep_2017} implemented its activation functions by detecting each optical signal, feeding them through a conventional digital computer to apply the nonlinearity, and then modulating new optical signals for the subsequent layer. 
Although this approach benefits from the flexibility of digital signal processing, conventional processors have a limited number of input and output channels, which make it challenging to scale this approach to very large matrix dimensions, which corresponds to a large number of optical inputs. 
Moreover, digitally applied nonlinearities add latency from the analog-to-digital conversion process and constrain the computational speed of the neural network to the same GHz-scale clock rates which ONNs seek to overcome. 
Thus, a hardware nonlinear optical activation, which doesn't require repeated bidirectional optical-electronic signal conversion, is of fundamental interest for making integrated ONNs a viable machine learning platform.

In this article, we propose an electro-optic architecture for synthesizing optical-to-optical nonlinearities which alleviates the issues discussed above. 
Our architecture features complete \textit{on}-\textit{off} contrast in signal transmission, a variety of nonlinear response curves, and a low activation threshold. 
Rather than using traditional optical nonlinearities, our scheme operates by measuring a small portion of the incoming optical signal power and using electro-optic modulators to modulate the original optical signal, without any reduction in operating bandwidth or computational speed. 
Additionally, our scheme allows for the possibility of performing additional nonlinear transformations on the signal using analog electrical components. 
Related electro-optical architectures for generating optical nonlinearities have been previously considered \cite{lentine_evolution_1993, majumdar_cavityenabled_2014, tait_silicon_2019}. 
In this work, we focus on the application of our architecture as an element-wise activation in a feedforward ONN, but the synthesis of low-threshold optical nonlinearities could be of broader interest to optical computing and information processing.

The remainder of this paper is organized as follows. 
First, we review the basic operating principles of ANNs and their integrated optical implementations in waveguide interferometer meshes. 
We then introduce our electro-optical activation function architecture, showing that it can be reprogrammed to synthesize a variety of nonlinear responses. 
Next, we discuss the performance of an ONN using this architecture by analyzing the scaling of power consumption, latency, processing speed, and footprint. 
We then draw an analogy between our proposed activation function and the optical Kerr effect. 
Finally, using numerical simulations, we demonstrate that our architecture leads to improved performance on two different machine learning tasks: (1) learning an N-input exclusive OR (XOR) logic function; (2) classifying images of handwritten numbers from the MNIST dataset.

\begin{figure*}
  \centering
  \includegraphics{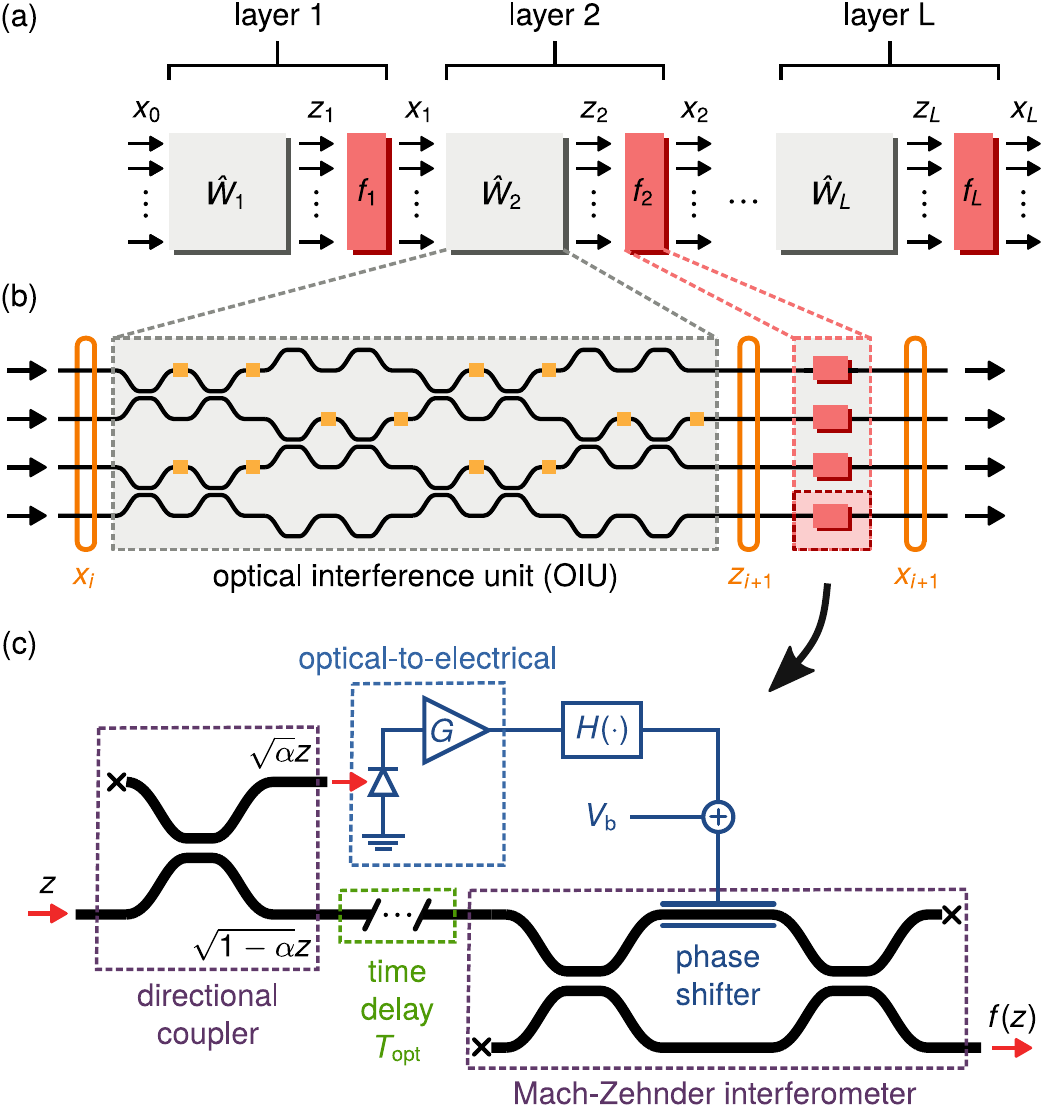}
  \caption{(a) Block diagram of a feedforward neural network of $L$ layers. 
  Each layer consists of a $\hat{W}_i$ block representing a linear matrix which multiplies vector inputs $x_{i-1}$. 
  The $f_i$ block in each layer represents an element-wise nonlinear activation function operating on vectors $z_i$ to produce outputs $x_{i}$.
  (b) Schematic of the optical interferometer mesh implementation of a single layer of the feedforward neural network. 
  (c) Schematic of the proposed optical-to-optical activation function which achieves a nonlinear response by converting a small portion of the optical input, $z$ into an electrical signal, and then intensity modulating the remaining portion of the original optical signal as it passes through an interferometer.}
  \label{fig:overview}
\end{figure*}

\section{Feedforward Optical Neural Networks} 
\label{sec:onn}

In this section, we briefly review the basics of feedforward artificial neural networks (ANNs) and describe their implementation in a reconfigurable optical circuit, as proposed in Ref. \citenum{shen_deep_2017}.  
As outlined in Fig. \ref{fig:overview}(a), an ANN is a function which accepts an input vector, $x_0$ and returns an output vector, $x_L$.  
This is accomplished in a layer-by-layer fashion, with each layer consisting of a linear matrix-vector multiplication followed by the application of an element-wise nonlinear function, or \textit{activation}, on the result.  
For a layer with index $i$, containing a weight matrix $\hat{W}_i$ and activation function $f_i(\cdot)$, its operation is described mathematically as
\begin{equation}
   x_i = f_i{\left( \hat{W}_i \cdot x_{i-1} \right)}
\end{equation}
for $i$ from 1 to $L$.

Before they are able to perform a given machine learning task, ANNs must be trained. 
The training process is typically accomplished by minimizing the prediction error of the ANN on a set of training examples, which come in the form of input and target output pairs. 
For a given ANN, a loss function is defined to quantify the difference between the target output and output predicted by the network.  
During training, this loss function is minimized with respect to tunable degrees of freedom, namely the elements of the weight matrix $\hat{W}_i$ within each layer. 
In general, although less common, it is also possible to train the parameters of the activation functions \cite{trentin_networks_2001}.

Optical hardware implementations of ANNs have been proposed in various forms over the past few decades.  
In this work, we focus on a recent demonstration in which the linear operations are implemented using an integrated optical circuit \cite{shen_deep_2017}.  
In this scheme, the information being processed by the network, $x_i$, is encoded into the modal amplitudes of the waveguides feeding the device and the matrix-vector multiplications are accomplished using meshes of integrated optical interferometers.  
In this case, training the network requires finding the optimal settings for the integrated optical phase shifters controlling the inteferometers, which may be found using an analytical model of the chip, or using \textit{in-situ} backpropagation techniques \cite{hughes_training_2018}.

In the next section, we present an approach for realizing the activation function, $f_i(\cdot)$, on-chip with a hybrid electro-optic circuit feeding an inteferometer.  
In Fig. \ref{fig:overview}(b), we show how this activation scheme fits into a single layer of an ONN and show the specific form of the activation in Fig. \ref{fig:overview}(c). 
We also give the specific mathematical form of this activation and analyze its performance in practical operation.

\section{Nonlinear Activation Function Architecture}
\label{sec:activation}

In this section, we describe our proposed nonlinear activation function architecture for optical neural networks, which implements an optical-to-optical nonlinearity by converting a small portion of the optical input power into an electrical voltage. 
The remaining portion of the original optical signal is phase- and amplitude-modulated by this voltage as it passes through an interferometer. 
For an input signal with amplitude $z$, the resulting nonlinear optical activation function, $f(z)$, is a result of the responses of the interferometer under modulation as well as the components in the electrical signal pathway.

A schematic of the architecture is shown in Fig. \ref{fig:overview}(c), where black and blue lines represent optical waveguides and electrical signal pathways, respectively. 
The input signal first enters a directional coupler which routes a portion, $\alpha$, of the input optical power to a photodetector. 
The photodetector is the first element of an optical-to-electrical conversion circuit, which is a standard component of high-speed optical receivers for converting an optical intensity into a voltage. 
In this work, we assume a normalization of the optical signal such that the total power in the input signal is given by $\vert z \vert^2$. 
The optical-to-electrical conversion process consists of the photodetector producing an electrical current, $I_{\text{pd}} = \mathfrak{R} \cdot \alpha \vert z \vert^2$, where $\mathfrak{R}$ is the photodetector responsivity, and a transimpedance amplifying stage, characterized by a gain $G$, converting this current into a voltage $V_G = G \cdot \mathfrak{R} \cdot \alpha \vert z \vert^2$. 
The output voltage of the optical-to-electrical conversion circuit then passes through a nonlinear signal conditioner with a transfer function, $H(\cdot)$. 
This component allows for the application of additional nonlinear functions to transform the voltage signal. 
Finally, the conditioned voltage signal, $H(V_G)$ is combined with a static bias voltage, $V_b$ to induce a phase shift of
\begin{equation}
    \Delta{\phi} = \frac{\pi}{V_\pi} \left[V_b + H{\left(G \mathfrak{R} \alpha \vert z \vert^2 \right)} \right] \label{eq:nlspm}
\end{equation}
for the optical signal routed through the lower port of the directional coupler. 
The parameter $V_\pi$ represents the voltage required to induce a phase shift of $\pi$ in the phase modulator. 
This phase shift, defined by Eq. \ref{eq:nlspm}, is a nonlinear self-phase modulation because it depends on the input signal intensity.

An optical delay line between the directional coupler and the Mach-Zehnder interferometer (MZI) is used to match the signal propagation delays in the optical and electrical pathways. 
This ensures that the nonlinear self-phase modulation defined by Eq. \ref{eq:nlspm} is applied at the same time that the optical signal which generated it passes through the phase modulator. 
For the circuit shown in Fig. \ref{fig:overview}(c), the optical delay is $\tau_{\text{opt}} = \tau_{\text{oe}} + \tau_{\text{nl}} + \tau_{\text{rc}}$, accounting for the contributions from the group delay of the optical-to-electrical conversion stage ($\tau_{\text{oe}}$), the delay associated with the nonlinear signal conditioner ($\tau_{\text{nl}}$), and the RC time constant of the phase modulator ($\tau_{\text{rc}}$). 

The nonlinear self-phase modulation achieved by the electric circuit is converted into a nonlinear amplitude response by the MZI, which has a transmission depending on $\Delta{\phi}$ as
\begin{equation}
    t_{\text{MZI}} = j \exp{\left( -j \frac{\Delta{\phi}}{2} \right)} \cos{\left( \frac{\Delta{\phi}}{2} \right)} \label{eq:Tmzi}.
\end{equation}
Depending on the configuration of the bias, $V_b$, a larger input optical signal amplitude causes either more or less power to be diverted away from the output port, resulting in a nonlinear self-intensity modulation. 
Combining the expression for the nonlinear self-phase modulation, given by Eq. \ref{eq:nlspm}, with the MZI transmission, given by Eq. \ref{eq:Tmzi}, the mathematical form of the activation function can be written explicitly as
\begin{multline}
   f{\left(z\right)} = j \sqrt{1-\alpha} \exp{\left( -j \frac{1}{2} \left[\phi_b + \pi\frac{H{\left(G \mathfrak{R} \alpha \vert z \vert^2 \right)}}{V_{\pi}} \right] \right)} \\ \cdot \cos{\left( \frac{1}{2} \left[\phi_b + \pi\frac{H{\left(G \mathfrak{R} \alpha \vert z \vert^2 \right)}}{V_{\pi}} \right] \right)} z, \label{eq:fx_general)}
\end{multline}
where the contribution to the phase shift from the bias voltage is
\begin{equation}
   \phi_b = \pi \frac{V_b}{V_\pi}. \label{eq:phi_b}
\end{equation}

\begin{figure}
  \centering
  \includegraphics{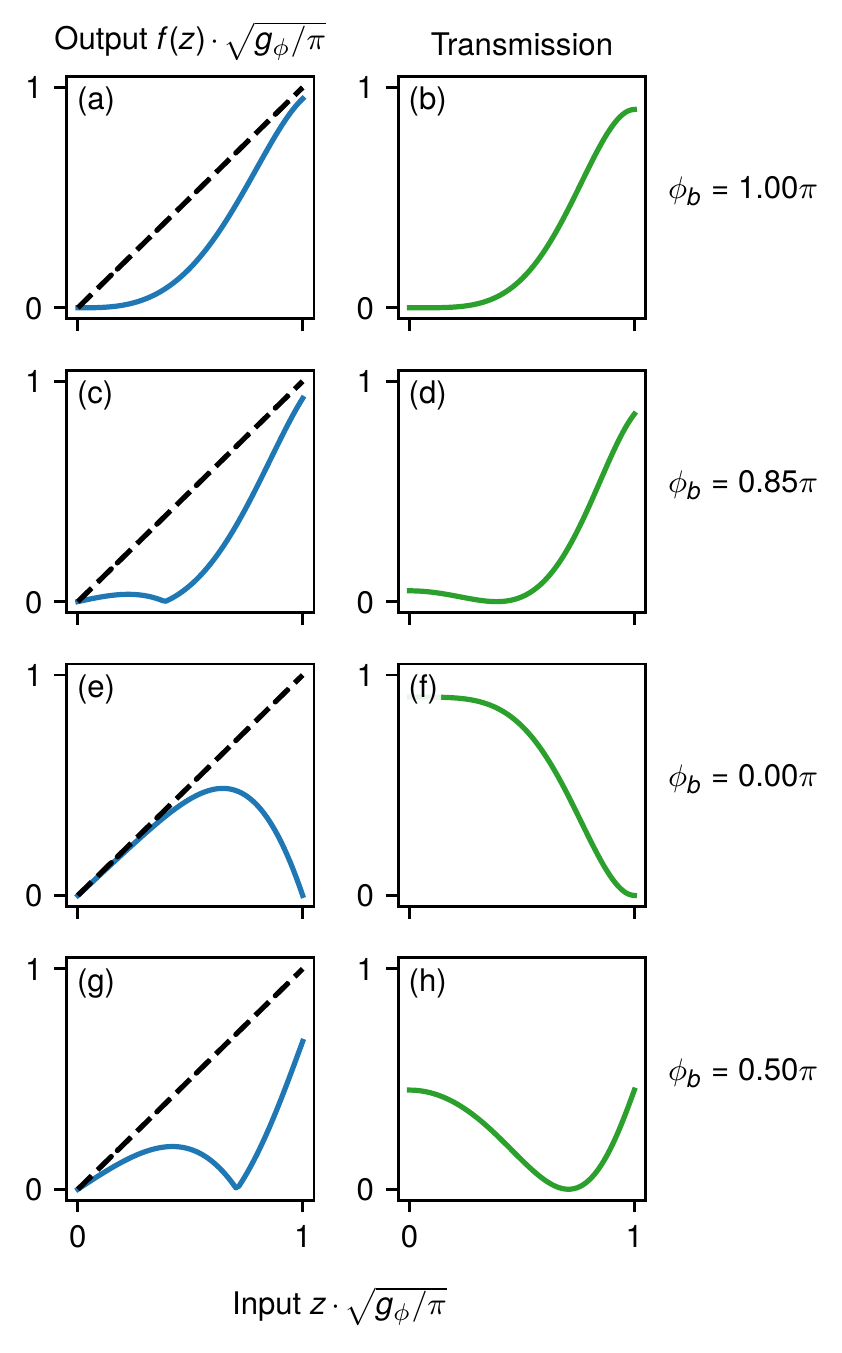}
  \caption{Activation function output amplitude (blue lines) and activation function transmission (green lines) as a function of input signal amplitude. 
  The input and output are normalized to the phase gain parameter, $g_\phi$.
  Panel pairs (a),(b) and (c),(d) correspond to a \texttt{ReLU}-like response, with a suppressed transmission for inputs with small amplitude and high transmission for inputs with large amplitude. 
  Panel pairs (e),(f) and (g),(h) correspond to a clipped response, with high transmission for inputs with small amplitude and reduced transmission for inputs with larger amplitude.}
  \label{fig:activation}
\end{figure}

For the remainder of this work, we focus on the case where no nonlinear signal conditioning is applied to the electrical signal pathway, i.e. $H{\left(V_G\right)} = V_G$. 
However, even with this simplification the activation function still exhibits a highly nonlinear response. 
We also neglect saturating effects in the OE conversion stage which can occur in either the photodetector or the amplifier. 
However, in practice, the nonlinear optical-to-optical transfer function could take advantage of these saturating effects. 

With the above simplifications, a more compact expression for the activation function response is
\begin{multline}
   f{\left(z\right)} = j \sqrt{1-\alpha} \exp{\left(-j \left[\frac{g_\phi \left\vert z \right\vert^2}{2} + \frac{\phi_b}{2} \right] \right)} \\ \cdot \cos{\left( \frac{g_\phi \left\vert z \right\vert^2}{2}  + \frac{\phi_b}{2} \right)} z, \label{eq:fx}
\end{multline}
where the phase gain parameter is defined as
\begin{equation}
   g_\phi = \pi \frac{ \alpha G \mathfrak{R} }{ V_{\pi} }. \label{eq:g}
\end{equation}
Equation \ref{eq:g} indicates that the amount of phase shift per unit input signal power can be increased via the gain and photodiode responsivity, or by converting a larger fraction of the optical power to the electrical domain. 
However, tapping out a larger fraction optical power also results in a larger linear loss, which is undesirable.

The electrical biasing of the activation phase shifter, represented by $V_b$, is an important degree of freedom for determining its nonlinear response. 
We consider a representative selection, consisting of four different responses, in Fig. \ref{fig:activation}. 
The left column of Fig. \ref{fig:activation} plots the output signal amplitude as a function of the input signal amplitude i.e. $\vert f(z) \vert$ in Eq. \ref{eq:fx}, while the right column plots the transmission coefficient i.e. $\vert f(z) \vert^2/\vert z \vert^2$, a quantity which is more commonly used in optics than machine learning. 
The first two rows of Fig. \ref{fig:activation}, corresponding to $\phi_b = 1.0\pi$ and $0.85\pi$, exhibit a response which is comparable to the \texttt{ReLU} activation function: transmission is low for small input values and high for large input values. 
For the bias of $\phi_b = 0.85\pi$, transmission at low input values is slightly increased with respect to the response where $\phi_b = 1.00\pi$. 
Unlike the ideal \texttt{ReLU} response, the activation at $\phi_b = 0.85\pi$ is not entirely monotonic because transmission first goes to zero before increasing. 
On the other hand, the responses shown in the bottom two rows of Fig. \ref{fig:activation}, corresponding to $\phi_b = 0.0\pi$ and $0.50\pi$, are quite different. 
These configurations demonstrate a saturating response in which the output is suppressed for higher input values but enhanced for lower input values. 
For all of the responses shown in Fig. \ref{fig:activation}, we have assumed $\alpha = 0.1$ which limits the maximum transmission to $1-\alpha = 0.9$.

A benefit of having electrical control over the activation response is that, in principle, its electrical bias can be connected to the same control circuitry which programs the linear interferometer meshes. 
In doing so, a single ONN hardware unit can then be reprogrammed to synthesize many different activation function responses. 
This opens up the possibility of heuristically selecting an activation function response, or directly optimizing the the activation bias using a training algorithm. 
This realization of a flexible optical-to-optical nonlinearity can allow ONNs to be applied to much broader classes of machine learning tasks.

We note that Fig. \ref{fig:activation} shows only the amplitude response of the activation function. 
In fact, all of these responses also introduce a nonlinear self-phase modulation to the output signal. 
If desired, this nonlinear self-phase modulation can be suppressed using a push-pull interferometer configuration in which the generated phase shift, $\Delta{\phi}$, is divided and applied with opposite sign to the top and bottom arms.

\section{Performance and Scalability}

In this section, we discuss the performance of an integrated ONN which uses meshes of integrated optical interferometers to perform matrix-vector multiplications and the electro-optic activation function, as shown in Fig. \ref{fig:overview}(b),(c). 
Here, we focus on characterizing how the power consumption, computational latency, physical footprint, and computational speed of the ONN scale with respect to the number of network layers, $L$ and the dimension of the input vector, $N$, assuming square matrices.
The system parameters used for this analysis are summarized in Table \ref{tab:parameters} and the figures of merit are summarized in Table \ref{tab:fom}.

\begin{table}
\caption{Summary of parameter values}
\setlength{\tabcolsep}{5pt}
\renewcommand{\arraystretch}{1.3}
\centering
\begin{tabular}{|l|c|}
\hline
\textbf{Parameter}                                              & \textbf{Value} \\
\hline
Modulator and detector rate                                     & 10 GHz \\
Photodetector responsivity ($\mathfrak{R}$)                     & 1 A/W \\
Optical-to-electrical circuit power consumption                 & 100 mW \\
Optical-to-electrical circuit group delay ($\tau_{\text{eo}}$)  & 100 ps \\
Phase modulator RC delay ($\tau_{\text{rc}}$)                   & 20 ps \\
Mesh MZI length $\left(D_{\text{MZI}}\right)$                   & 100 $\mu$m \\
Mesh MZI height $\left(H_{\text{MZI}}\right)$                   & 60  $\mu$m \\
Waveguide effective index ($n_{\text{eff}}$)                    & 3.5 \\
\hline
\end{tabular}
\label{tab:parameters}
\end{table}

\begin{table*}
\caption{Summary of per-layer optical neural network performance using the electro-optic activation function}
\setlength{\tabcolsep}{10pt}
\renewcommand{\arraystretch}{1.3}
\centering
\begin{tabular}{|l|ll|lll|}
\hline
                  & \multicolumn{2}{c|}{\textbf{Scaling}}  & \multicolumn{3}{c|}{\textbf{Per-layer figures of merit}} \\
                  & \textbf{Mesh}   & \textbf{Activation}                 & $N=4$                       & $N=10$                            & $N=100$                           \\
\hline
\textbf{Power consumption}$^*$ &  & $L N$                       & 0.4 W                       & 1 W                         & 10 W                        \\
\textbf{Latency}           & $L N$   & $L$                        & 125 ps                      & 132 ps                      & 237 ps                      \\
\textbf{Footprint}         & $L N^2$ & $L N$  & 2.5 mm$^2$ &  6.6 mm$^2$ & 120.0 mm$^2$                               \\
\textbf{Speed}      & $L N^2$ &                            & $1.6\times 10^{11}$ MAC/s   & $1 \times 10^{12}$ MAC/s  & $1 \times 10^{14}$ MAC/s  \\
\hline
\textbf{Efficiency}$^*$    & \multicolumn{2}{c|}{$N^{-1}$}                              & 2.5 pJ/MAC                        & 1 pJ/MAC                          & 100 fJ/MAC                        \\
\hline
\end{tabular}
\\
\vspace{3pt}
\footnotesize{$^*$Assuming no power consumption in the interferometer mesh phase shifters}
\label{tab:fom}
\end{table*}

\subsection{Power consumption}

\begin{figure}
  \centering
  \includegraphics{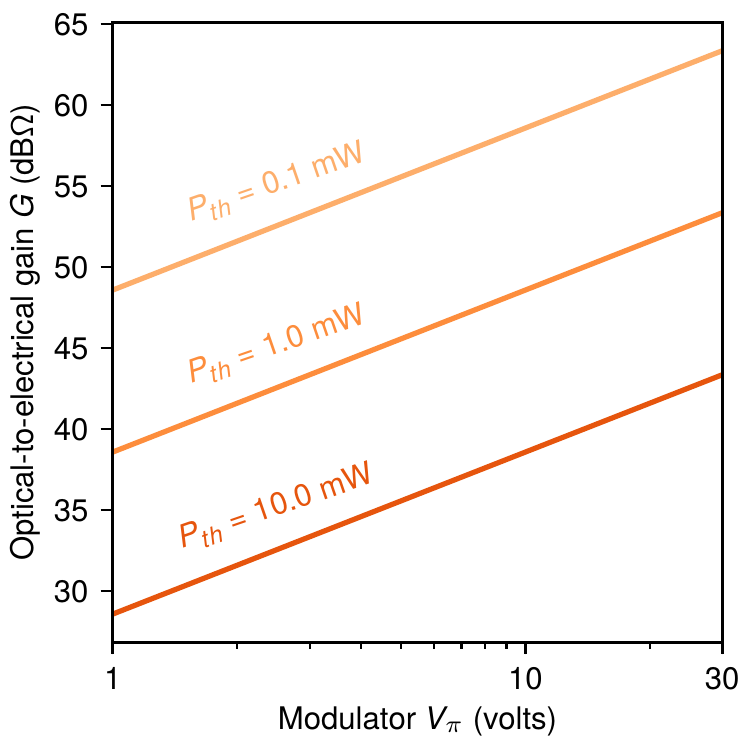}
  \caption{Contours of constant activation threshold as a function of the optical-to-electrical gain and the modulator $V_\pi$ of the activation function shown in Fig. \ref{fig:overview}(c) with a photodetector responsivity $\mathfrak{R} $ = 1.0 A/W.}
  \label{fig:threshold}
\end{figure}

The power consumption of the ONN, as shown in Fig. \ref{fig:overview}(b), consists of contributions from (1) the programmable phase shifters inside the interferometer mesh, (2) the optical source supplying the input vectors, $x_0$, and (3) the active components of the activation function such as the amplifier and photodetector. 
In principle, the contribution from (1) can be made negligible by using phase change materials or ultra-low power MEMS phase shifters.
Therefore, in this section we focus only on contributions (2) and (3) which pertain to the activation function.

To quantify the power consumption, we first consider the minimum input optical power to a \textit{single} activation that triggers a nonlinear response. 
We refer to this as the activation function threshold, which is mathematically defined as
\begin{equation}
    P_{\text{th}} = \frac{\Delta{\phi}\vert_{\delta{T}=0.5}}{g_\phi} = \frac{V_\pi}{\pi \alpha G \mathfrak{R}} \cdot \Delta{\phi}\vert_{\delta{T}=0.5} \label{eq:Pth},
\end{equation}
where $\Delta{\phi}\vert_{\delta{T}=0.5}$ is the is phase shift necessary to generate a 50\% change in the power transmission with respect to the transmission with null input for a given $\phi_b$. 
This threshold corresponds to $z\sqrt{g_\phi/\pi} = 0.73$ in Fig. \ref{fig:activation}(b), to $z\sqrt{g_\phi/\pi}=0.85$ in Fig. \ref{fig:activation}(d), to $z\sqrt{g_\phi/\pi}=0.73$ in Fig. \ref{fig:activation}(f), and to $z\sqrt{g_\phi/\pi}=0.70$ in Fig. \ref{fig:activation}(h).
In general, a lower activation threshold will result in a lower optical power required at the ONN input, $\vert x_0 \vert^2$. 
According to Eq. \ref{eq:Pth}, the activation threshold can be reduced via a small $V_\pi$ and a large optical-to-electrical conversion gain, $G\mathfrak{R} \sim 1.0$ V/mW.
The relationship between $G$ and $V_\pi$ for activation thresholds of 0.1 mW, 1.0 mW, and 10.0 mW is shown in Fig. \ref{fig:threshold} for a fixed $\mathfrak{R}$ = 1 A/W. 
Additionally, in Fig. \ref{fig:threshold} we conservatively assume $\phi_b = \pi$ which has the highest threshold of the activation function biases shown in Fig. \ref{fig:activation}.

If we take the lowest activation threshold of 0.1 mW in Fig. \ref{fig:threshold}, the optical source to the ONN would then need to supply $N \cdot 0.1$ mW of optical power. 
The power consumption of integrated optical receiver amplifiers varies considerably, ranging from as low as 10 mW to as high as 150 mW \cite{ahmed_100_2014, settaluri_first_2017, nozaki_amplifier-free_2018}, depending on a variety of factors which are beyond the scope of this article.
Therefore, a conservative estimate of the power consumption from the optical-to-electrical conversion circuits in all activations is $L \cdot N \cdot 100$ mW.
For an ONN with $N = 100$, the power consumption per layer from the activation function would be 10 W and would require a total optical input power of $N \cdot P_{\text{th}} = 100 \cdot 0.1$ mW = 10 mW. Thus, the total power consumption of the ONN is dominated by the activation function electronics.

\subsection{Latency}
For the feedforward neural network architecture shown in Fig. \ref{fig:overview}(a), the latency is defined by the elapsed time between supplying an input vector, $x_0$ and detecting its corresponding prediction vector, $x_L$. 
In an integrated ONN, as implemented in Fig. \ref{fig:overview}(b), this delay is simply the travel time for an optical pulse through all $L$-layers. 
Following Ref. \citenum{shen_deep_2017}, the propagation distance in a square interferometer mesh is $D_{W} = N \cdot D_{\text{MZI}}$, where $D_{\text{MZI}}$ is the length of each MZI within the mesh. 
In the nonlinear activation layer, the propagation length will be dominated by the delay line required to match the optical and electrical delays, and is given by
\begin{equation}
    D_{f} = \left( \tau_{\text{oe}} + \tau_{\text{nl}} + \tau_{\text{rc}} \right) \cdot v_g,
\end{equation}
where the group velocity $v_g = c_0 / n_{\text{eff}}$ is the speed of optical pulses in the waveguide. Therefore, 
\begin{equation}
    \text{latency} = \underbrace{ L \cdot N \cdot D_{\text{MZI}} \cdot {v_g}^{-1} }_{\text{Interferometer mesh}} + \underbrace{ L \cdot \left( \tau_{\text{oe}} + \tau_{\text{nl}} + \tau_{\text{rc}} \right)}_{\text{Activation function}}. \label{eq:latency}
\end{equation}
Equation \ref{eq:latency} indicates that the latency contribution from the interferometer mesh scales with the product $LN$, which is the same scaling as predicted in Ref. \citenum{shen_deep_2017}. 
On the other hand, the activation function adds to the latency independently of $N$ because each activation circuit is applied in parallel to all $N$-vector elements.

For concreteness, we assume $D_{\text{MZI}} = 100\ \mu\text{m}$ and $n_{\text{eff}} = 3.5$. 
Following our assumption in the previous section of using no nonlinear electrical signal conditioner in the activation function, $\tau_{\text{nl}}$ = 0 ps. 
Typical group delays for integrated transimpedance amplifiers used in optical receivers can range from $\tau_{\text{oe}} \approx$ 10 to 100 ps. 
Moreover, assuming an RC-limited phase modulator speed of 50 GHz yields $\tau_{\text{rc}} \approx 20$ ps. Therefore, if we assume a conservative value of $\tau_{\text{oe}}$ = 100 ps, a network dimension of $N \approx 100$ would have a latency of $237$ ps per layer, with equal contributions from the mesh and the activation function. 
For a ten layer network ($L = 10$) the total latency would be approximately 2.4 ns, still orders of magnitude lower than the latency typically associated with GPUs.

\subsection{Physical footprint}
The physical footprint of the ONN consists of the space taken up by both the linear interferometer mesh and the optical and electrical components of the activation function. 
Neglecting the electrical control lines for tuning each MZI, the total footprint of the ONN is
\begin{equation}
    A = \underbrace{ L\cdot N^2 \cdot A_{\text{MZI}} }_{\text{Interferometer mesh}} + \underbrace{ L\cdot N \cdot A_{\text{f}} }_{\text{Activation function}}, \label{eq:area}
\end{equation}
where $A_{\text{MZI}} = D_{\text{MZI}} \cdot H_{\text{MZI}}$ is the area of a single MZI element in the mesh and $A_{\text{f}} = D_{\text{f}} \cdot H_{\text{f}}$ is the area of a single activation function.

In the direction of propagation, $D_{\text{f}}$ is dominated by the waveguide optical delay line required to match the delay of the electrical signal pathway.
Based on the previous discussion of the activation function's latency, $\tau_{\text{opt}} = 120$ ps corresponds to a total waveguide length of $D_{\text{f}}$ $\approx$ 1 cm.
For simplicity, we assume this delay is achieved using a straight waveguide, which results in a large footprint but with optical losses that can be very low.
For example, in silicon waveguides losses below 0.5 dB/cm have been experimentally demonstrated \cite{selvaraja_highly_2014}.
In principle, incorporating waveguide bends or resonant optical elements could significantly reduce the activation function's footprint. 
For example, coupled micro ring arrays have experimentally achieved group delays of 135 ps over a bandwidth of 10 GHz in a 0.03 mm $\times$ 0.25 mm footprint \cite{cardenas_widebandwidth_2010}.

Transverse to the direction of propagation, the activation function footprint will be dominated by the electronic components of the optical-to-electrical conversion circuit.
In principle, compact waveguide photodetectors and modulators can be utilized.
However, the components of the transimpedance amplifier may be challenging to integrate in the area available between neighboring output waveguides of the interferometer mesh.
One possibility towards achieving a fully integrated opto-electronic ONN would be to use so-called \textit{amplifier-free} optical receivers \cite{nozaki_amplifier-free_2018}, where ultra-low capacitance detectors provide high-speed opto-electronic conversion.
Similarly to the experimental demonstration in Ref. \citenum{nozaki_femtofarad_2019}, the amplifier-free receiver could be integrated directly with a high efficiency (e.g. effectively a low $V_\pi$) electro-optic modulator.
Compact electro-absorption modulators could also be utilized.
In addition to achieving a compact footprint, operating without an amplifier would also result in an order of magnitude reduction in both power consumption and latency, with the later reducing the required length of the optical delay line and thus the footprint.

For the purposes of our analysis, we assume no integration of the electronic transimpedance amplifier and, therefore, that the on-chip components of the activation function fit within the height of each interferometer mesh row, $D_{\text{f}} \le D_{\text{MZI}}$ = 60 $\mu$m. Under this assumption and following the scaling in Eq. \ref{eq:area}, the total footprint of a single ONN layer of dimension $N=10$ would be 11.0 mm $\times$ 0.6 mm. Interestingly, following the latency discussion in the previous section, a single ONN layer of dimension $N=100$ would have a footprint of 20.0 mm $\times$ 6.0 mm, with equal contribution from the activation function and from the mesh.

\subsection{Speed}
The speed, or computational capacity, of the ONN, as shown in Fig. \ref{fig:overview}(a), is determined by the number of input vectors, $x_0$ that can be processed per unit time. 
Here, we argue that although our activation function is not fully optical, it results in no speed degradation compared to a linear ONN consisting of only interferometer meshes.

The reason for this is that a fully integrated ONN would also include high-speed modulators and detectors on-chip to perform fast modulation and detection of sequences of $x_0$ vectors and $x_L$ vectors, respectively. 
We therefore argue that the same high-speed detector and modulator elements could also be integrated between the linear network layers to provide the optical-electrical and electrical-optical transduction for the activation function. 
State of the art integrated transimpedance amplifiers can already operate at speeds comparable to the optical modulator and detector rates, which are on the order of 50 - 100 GHz \cite{yu_low-noise_2012, ahmed_100_2014}, and thus would not be a limiting factor in the speed of our architecture. 

To perform a matrix-vector multiplication on a conventional CPU requires $N^2$ multiply-accumulate (MAC) operations, each consisting of a single multiplication and a single addition.
Therefore, assuming a photodetector and modulator rate of 10 GHz means that an ONN can effectively perform $N^2 \cdot L \cdot 10^{10}$ MAC/sec. 
This means that one layer of an ONN with dimension $N=10$ would effectively perform $10^{12}$ MAC/sec. 
Increasing the input dimension to $N = 100$ would then scale the performance of the ONN to $10^{14}$ MAC/sec per layer. 
This is two orders of magnitude greater than the peak performance obtainable with modern GPUs, which typically have performance on the order of $10^{12}$ floating point operations/sec (FLOPS).
Because the power consumption of the ONN scales as $LN$ (assuming passive phase shifters in the mesh) and the speed scales as $LN^2$, the energy per operation is minimized for large $N$ (Table \ref{tab:fom}).
Thus, for large ONNs the power consumption associated with the electro-optic conversion in the activation function can be amortized over the parallelized operation of the linear mesh.

We note that the activation function circuit shown in Fig. \ref{fig:overview}(c) can be modified to remove the matched optical delay line by using very long optical pulses. 
This modification may be advantageous for reducing the footprint of the activation and would result in $\tau_\text{opt} \ll \tau_\text{ele}$. 
However, this results in a reduction of the ONN speed, which would then be limited by the combined activation delay of all $L$ nonlinear layers in the network, $\sim \left( L \cdot \tau_{\text{ele}} \right)^{-1}$.

\section{Comparison with the Kerr Effect}

\begin{figure}
  \centering
  \includegraphics{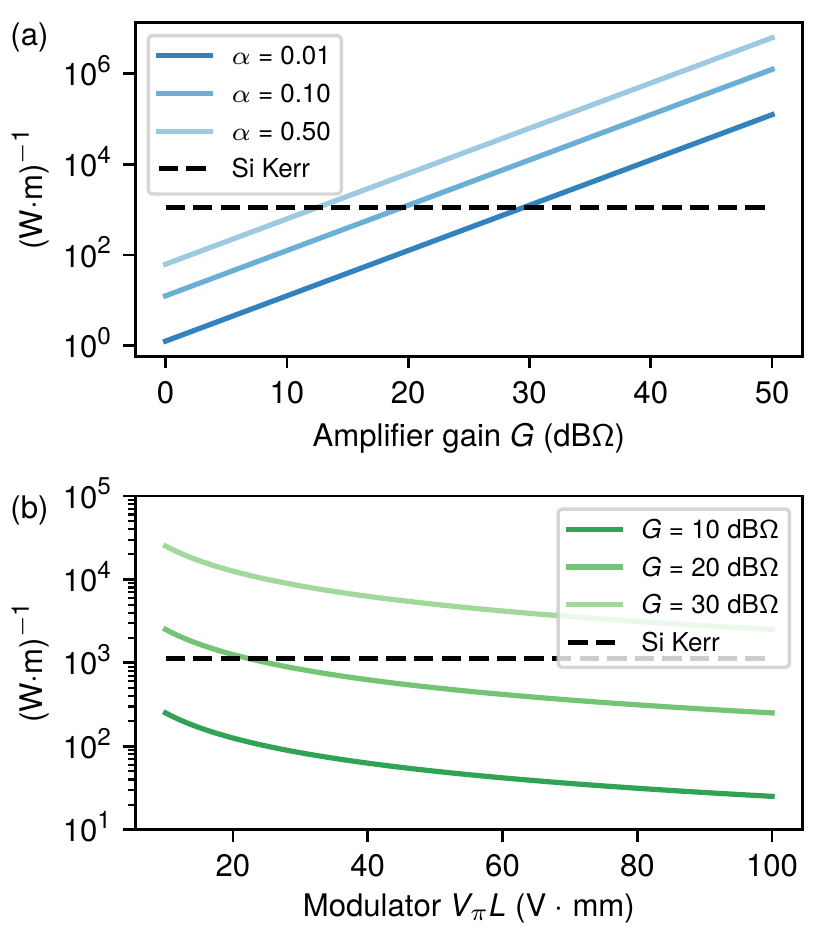}
  \caption{Nonlinear parameter $\Gamma_{\text{EO}}$ for the electro-optic activation as a function of (a) gain, $G$, for $\alpha$ = 0.50, 0.10, and 0.01 and (b) modulator $V_\pi L$. 
  The nonlinear parameter associated with the optical Kerr effect, $\Gamma_{\text{Kerr}}$ in a Silicon waveguide of cross sectional area $A = 0.05$ $\mu$m$^2$ corresponds to the black dotted line.}
  \label{fig:gamma}
\end{figure}

All-optical nonlinearities such as bistability and saturable absorption have been previously considered as potential activation functions in ONNs \cite{abu-mostafa_optical_1987,shen_supplementary_2017}.
An alternative implementation of the activation function in Fig. \ref{fig:overview}(c) could consist of a nonlinear MZI, with one of its arms having a material with Kerr nonlinear optical response.
The Kerr effect is a third-order optical nonlinearity which generates a change in the refractive index, and thus a nonlinear phase shift, which is proportional to the input pulse intensity. 
In this section we compare the electro-optic activation function introduced in the previous section [Fig. \ref{fig:overview}(c)] to such an alternative all-optical activation function using the Kerr effect, highlighting how the electro-optic activation can achieve a lower activation threshold.

Unlike the electro-optic activation function, the Kerr effect is lossless and has no latency because it arises from a nonlinear material response, rather than a feedforward circuit.
A standard figure of merit for quantifying the strength of the Kerr effect in a waveguide is through the amount of nonlinear phase shift generated per unit input power per unit waveguide length. 
This is given mathematically by the expression
\begin{equation}
	\Gamma_{\text{Kerr}} = \frac{2\pi}{\lambda_0}\frac{n_2}{A},
    \label{eqn:gamma_kerr}
\end{equation}
where $n_2$ is the nonlinear refractive index of the material and $A$ is the effective mode area. 
$\Gamma_{\text{Kerr}}$ ranges from 100 (W$\cdot$m)$^{-1}$ in chalcogenide to 350 (W$\cdot$m)$^{-1}$ in silicon \cite{koos_nonlinear_2007}. 
An equivalent figure of merit for the electro-optic feedforward scheme can be mathematically defined as
\begin{equation}
    \Gamma_{\text{EO}} = \pi \frac{\alpha \mathfrak{R} G}{V_\pi L},
    \label{eqn:gamma_eo}
\end{equation}
where $V_\pi L$ is the phase modulator figure of merit.
The figures of merit described in Eqs. \ref{eqn:gamma_kerr}-\ref{eqn:gamma_eo} can be represented as an activation threshold (Eq. \ref{eq:Pth}) via the relationship $P_{\text{th}} = \frac{\Delta{\phi}\vert_{\delta{T}=0.5}}{\Gamma L}$, for a given waveguide length, $L$ where the electro-optic phase shift or nonlinear Kerr effect take place.

A comparison of Eq. \ref{eqn:gamma_kerr} and Eq. \ref{eqn:gamma_eo} indicates that while the strength of the Kerr effect is largely fixed by waveguide design and material choice, the electro-optic scheme has several degrees of freedom which allow it to potentially achieve a stronger nonlinear response. The first design parameter is the amount of power tapped off to the photodetector, which can be increased to generate a larger voltage at the phase modulator. 
However, increasing $\alpha$ also increases the \textit{linear} signal loss through the activation which does not contribute to the nonlinear mapping between the input and output of the ONN. 
Therefore, $\alpha$ should be minimized as long as the optical power routed to the photodetector is large enough to be above the noise equivalent power level.

On the other hand, the product $\mathfrak{R}G$ determines the conversion efficiency of the detected optical power into an electrical voltage. 
Fig. \ref{fig:gamma}(a) compares the nonlinearity strength of the electro-optic activation (blue lines) to that of an implementation using the Kerr effect in silicon (black dashed line) for several values of $\alpha$, as a function of $G$. 
The responsivity is fixed at $\mathfrak{R} = 1.0$ A/W. 
We observe that tapping out 10\% of the optical power requires a gain of 20 dB$\Omega$ to achieve a nonlinear phase shift equivalent threshold to that of a silicon waveguide where $A$ = 0.05 $\mu$m$^2$ for the same amount of input optical power. 
Tapping out only 1\% of the optical power requires an additional 10 dB$\Omega$ of gain to maintain this equivalence. 
We note that the gain range considered in Fig. \ref{fig:gamma}(a) is well within the regime of what has been demonstrated in integrated transimpedance amplifiers for optical receivers \cite{ahmed_100_2014, settaluri_first_2017, nozaki_amplifier-free_2018}. 
In fact, many of these systems have demonstrated much higher gain. 
In Fig. \ref{fig:gamma}(a), the phase modulator $V_\pi L$ was fixed at 20 V$\cdot$mm. 
However, because a lower $V_\pi L$ translates into an increased phase shift for a given applied voltage, this parameter can also be used to enhance the nonlinearity. 
Fig. \ref{fig:gamma}(b) demonstrates the effect of changing the $V_\pi L$ for several values of of $G$, again, with a fixed responsivity $\mathfrak{R} = 1.0$ A/W. 
This demonstrates that with a reasonable level of gain and phase modulator performance, the electro-optic activation function can trade off an increase in latency for a significantly lower optical activation threshold than the Kerr effect.

\section{Machine Learning Tasks}
\label{sec:tasks}

In this section, we apply the electro-optic activation function introduced above to several machine learning tasks. 
In Sec. \ref{sec:xor}, we simulate training an ONN to implement an exclusive-OR (XOR) logical operation. 
The network is modeled using \texttt{neuroptica} \cite{noauthor_neuroptica_nodate}, a custom ONN simulator written in Python, which trains the simulated networks only from physically measurable field quantities using the on-chip backpropagation algorithm introduced in Ref. \citenum{hughes_training_2018}. 
In Sec. \ref{sec:mnist}, we consider the more complex task of using an ONN to classify handwritten digits from the Modified NIST (MNIST) dataset, which we model using the \texttt{neurophox} \cite{_neurophox_, pai_matrix_2019} package and \texttt{tensorflow} \cite{tensorflow2015-whitepaper}, which computes gradients using automatic differentiation.  
In both cases, we model the values in the network as complex-valued quantities and represent the interferometer meshes as unitary matrices parameterized by phase shifters.

\subsection{Exclusive-OR Logic Function} 
\label{sec:xor}

\begin{figure}
  \centering
  \includegraphics{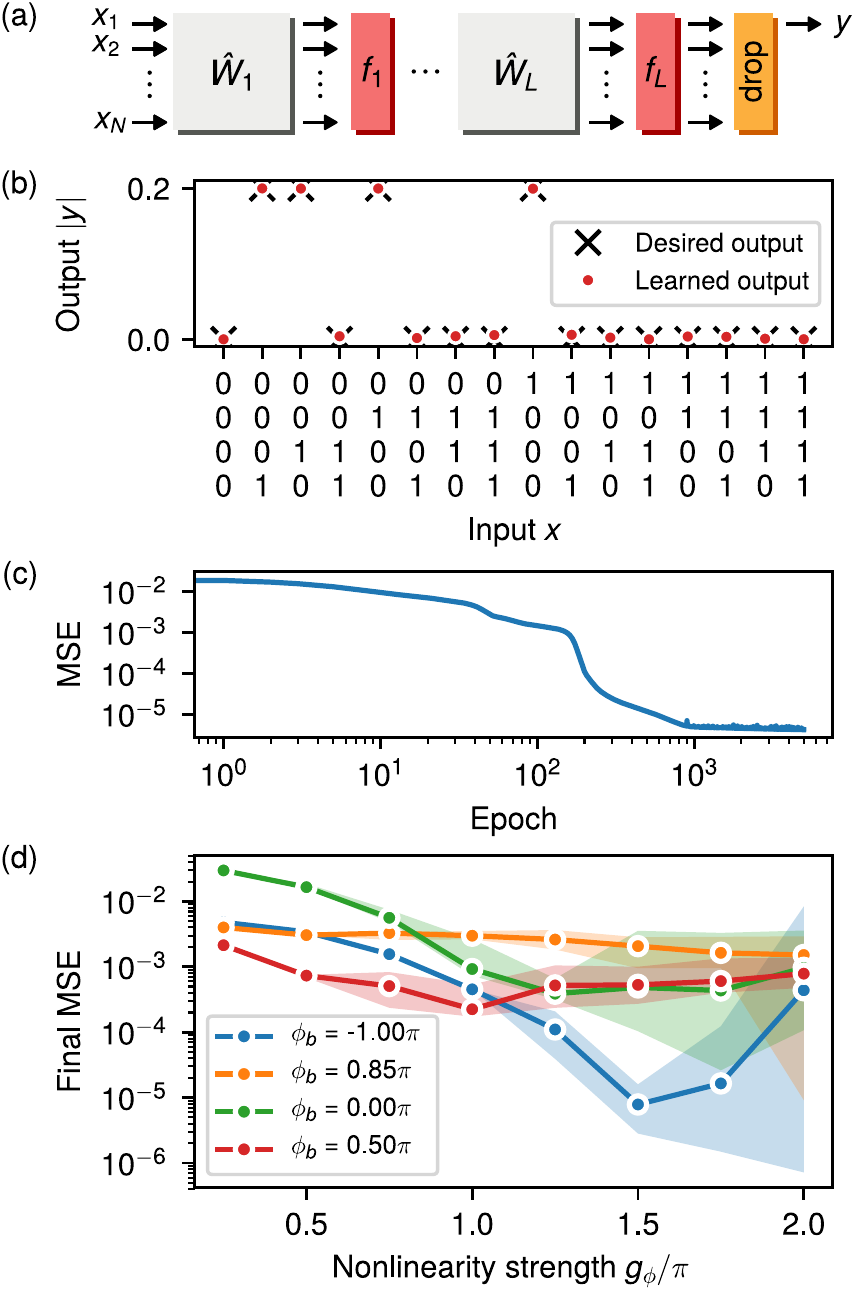}
  \caption{(a) Architecture of an $L$-layer ONN used to implement an $N$-input XOR logic function. 
  (b) Red dots indicate the learned input-output relationship of the XOR for $N=4$ on an 2-layer ONN. 
  Electro-optic activation functions are configured with gain $g = 1.75\pi$ and biasing phase $\phi_b = \pi$. 
  (c) Mean squared error (MSE) versus training epoch. 
  (d) Final MSE after 5000 epochs averaged over 20 independent training runs vs activation function gain. 
  Different lines correspond to the responses shown in Fig. \ref{fig:activation}, with $\phi_b = 1.00\pi$, $0.85\pi$, $0.00\pi$, and $0.50\pi$. 
  Shaded regions correspond to the range (minimum and maximum) final MSE from the 20 training runs.}
  \label{fig:xor}
\end{figure}

An exclusive-OR (XOR) is a logic function which takes two inputs and produces a single output. 
The output is \textit{high} if only one of the two inputs is \textit{high}, and \textit{low} for all other possible input combinations. 
In this example, we consider a multi-input XOR which takes $N$ input values, given by $x_1 \ldots x_N$, and produces a single output value, $y$. 
The input-output relationship of the multi-input XOR function is a generalization of the two-input XOR. 
For example, defining logical \textit{high} and \textit{low} values as 1 and 0, respectively, a four-input XOR has an output table indicated the desired values in Fig. \ref{fig:xor}(b). 
We select this task for the ONN to learn because it requires a non-trivial level of nonlinearity, meaning that it could not be implemented in an ONN consisting of only linear interferometer meshes.

The architecture of the ONN used to learn the XOR is shown schematically in Fig. \ref{fig:xor}(a). 
The network consists of $L$ layers, with each layer constructed from an $N \times N$ unitary interferometer mesh followed by an array of $N$ parallel electro-optic activation functions, with each element corresponding to the circuit in Fig. \ref{fig:overview}(c). 
After the final layer, the lower $N-1$ outputs are dropped to produce a single output value which corresponds to $y$. 
Unlike the ideal XOR input-output relationship described above, for the XOR task learned by the ONN we normalize the input vectors such that they always have an $L_2$ norm of 1. 
This constraint is equivalent to enforcing a constant input power to the network. 
Additionally, because the activation function causes the optical power level to be attenuated at each layer, we take the \textit{high} output state to be a value of 0.2, as shown in Fig. \ref{fig:overview}(b).
The \textit{low} output remains at a value of 0.0. 
An alternative to using a smaller amplitude for the output high state would be to add additional ports with fixed power biases to increase the total input power to the network, similarly to the XOR demonstrated in Ref. \citenum{hughes_training_2018}.

In Fig. \ref{fig:xor}(b) we show the four-input XOR input-output relationship which was learned by a two-layer ONN. 
The electro-optic activation functions were configured to have a gain of $g = 1.75\pi$ and biasing phase of $\phi_b = \pi$. 
This biasing phase configuration corresponds to the \texttt{ReLU}-like response shown in Fig. \ref{fig:activation}(a). 
The black markers indicate the desired output values while the red circles indicate the output learned by the two-layer ONN. 
Fig. \ref{fig:xor}(b) indicates excellent agreement between the learned output and the desired output. 
The evolution of the mean squared error (MSE) between the ONN output and the desired output during training confirms this agreement, as shown in Fig. \ref{fig:xor}(c), with a final MSE below $10^{-5}$.

To train the ONN, a total of $2^N = 16$ training examples were used, corresponding to all possible binary input combinations along the x-axis of Fig. \ref{fig:xor}(b). 
All 16 training examples were fed through the network in a batch to calculate the mean squared error (MSE) loss function. 
The gradient of the loss function with respect to each phase shifter was computed by backpropagating the error signal through the network to calculate the loss sensitivity at each phase shifter \cite{hughes_training_2018}.
The above steps were repeated until the MSE converged, as shown in Fig. \ref{fig:xor}(c).
Only the phase shifter parameters were optimized by the training algorithm, while all parameters of the activation function were unchanged.

To demonstrate that the nonlinearity provided by the electro-optic activation function is essential for the ONN to successfully learn the XOR, in Fig. \ref{fig:xor}(d) we plot the final MSE after 5000 training epochs, averaged over 20 independent training runs, as a function of the activation function gain, $g_\phi$. 
The shaded regions indicates the minimum and maximum range of the final MSE over the 20 training runs. 
The four lines shown in Fig. \ref{fig:xor}(d) correspond to the four activation function bias configurations shown in Fig. \ref{fig:activation}.

For the blue curve in Fig. \ref{fig:xor}(d), which corresponds to the ReLU-like activation, we observe a clear improvement in the final MSE with an increase in the nonlinearity strength. 
We also observe that for very high nonlinearity, above $g_\phi = 1.5\pi$, the range between the minimum and maximum final MSE broadens and the mean final MSE increases. 
However, the best case (minimum) final MSE continues to decrease, as indicated by the lower border of the shaded blue region. 
This trend indicates that although increasing nonlinearity improves the ONN's ability to learn the XOR function, very high levels of nonlinearity may also prevent the training algorithm from converging.

A trend of decreasing MSE with increasing nonlinearity is also observed for the activation corresponding to the green curve in Fig. \ref{fig:xor}(d). 
However, the range of MSE values begins to broaden at a lower value of $g_\phi = 1.0\pi$. 
Such broadening may be a result of the changing slope in the activation function output, as shown in Fig. \ref{fig:activation}(e). 
For the activation functions corresponding to the red and orange curves in Fig. \ref{fig:xor}(d), the final MSE decreases somewhat with an increase in $g_\phi$, but generally remains much higher than the other two activation function responses. 
We conclude that these two responses are not as well suited for learning the XOR function. 
Overall, these results demonstrate that the flexibility of our architecture to achieve specific forms of nonlinear activation functions is important for the successful operation of an ONN.

\subsection{Handwritten Digit Classification}
\label{sec:mnist}

\begin{figure*}[t]
  \centering
  \includegraphics{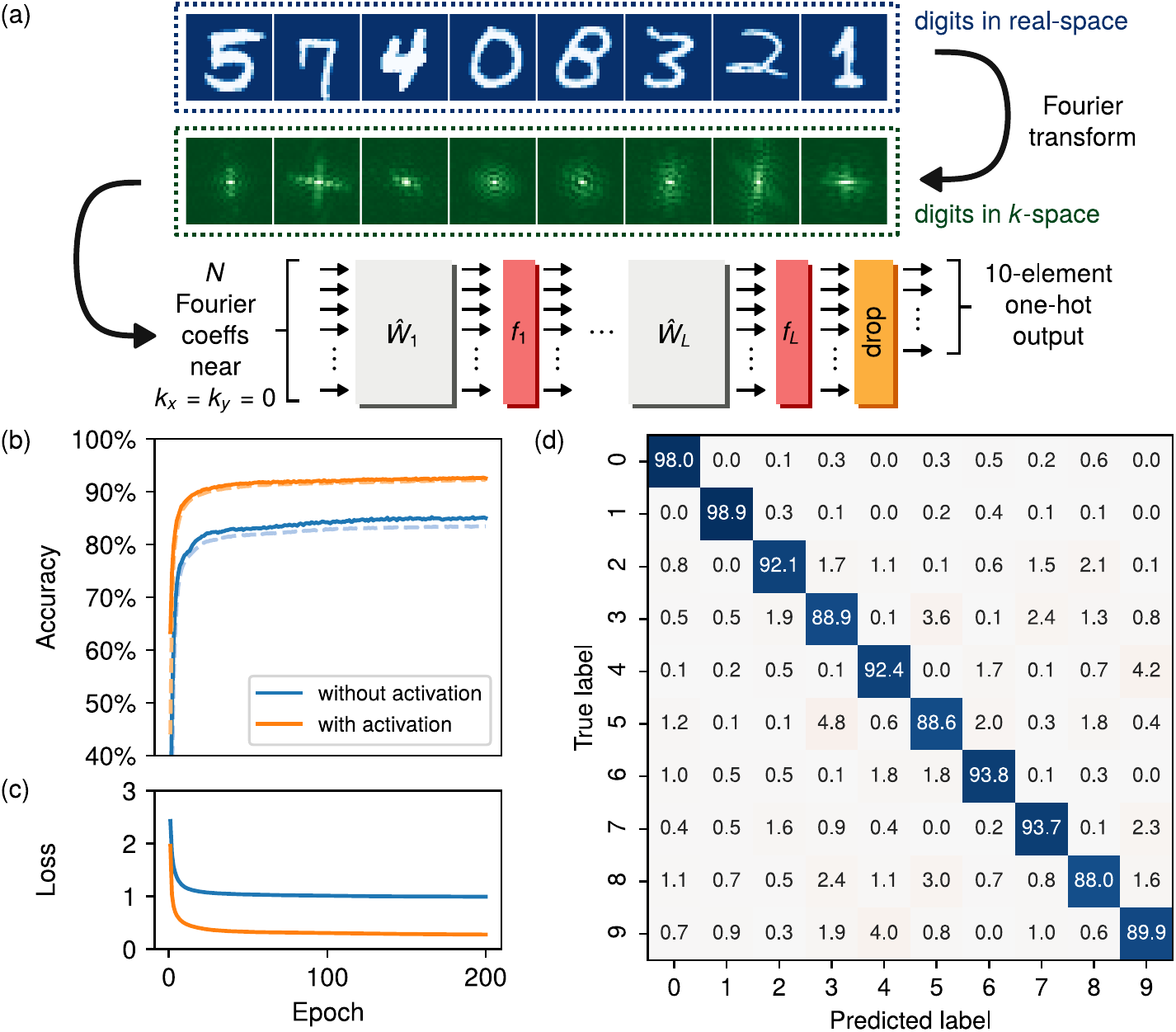}
  \caption{(a) Schematic of an optical image recognition setup based on an ONN. 
  Images of handwritten numbers from the MNIST database are preprocessed by converting from real-space to $k$-space and selecting $N$ Fourier coefficients associated with the smallest magnitude $k$-vectors. 
  (b) Test accuracy (solid lines) and training accuracy (dashed lines) during training for a two layer ONN without activation functions (blue) and with activation functions (orange). 
  $N=16$ Fourier components were used as inputs to the ONN and each vector was normalized such that its $L_2$ norm is unity. 
  The activation function parameters were $g_\phi = 0.05\pi$ and $\phi_b = 1.00\pi$. 
  (c) Cross entropy loss during training. 
  (d) Confusion matrix, specified in percentage, for the trained ONN with the electro-optic activation function.}
  \label{fig:mnist}
\end{figure*}

The second task we consider for demonstrating the activation function is classifying images of handwritten digits from the MNIST dataset, which has become a standard benchmark problem for ANNs \cite{lecun_gradient-based_1998}. 
The dataset consists of 70,000 grayscale 28$\times$28 pixel images of handwritten digits between 0 and 9. 
Several representative images from the dataset are shown in Fig. \ref{fig:mnist}(a).

To reduce the number of input parameters, and hence the size of the neural network, we use a preprocessing step to convert the images into a Fourier-space representation. 
Specifically, we compute the 2D Fourier transform of the images which is defined mathematically as $c{\left(k_x, k_y\right)} = \sum_{m, n} e^{j k_x m + j k_y n} g{\left(m, n\right)}$, where $g{\left(m, n\right)}$ is the gray scale value of the pixel at location $\left(m, n\right)$ within the image. 
The amplitudes of the Fourier coefficients $c{\left(k_x, k_y\right)}$ are shown below their corresponding images in Fig. \ref{fig:mnist}(a). 
These coefficients are generally complex-valued, but because the real-space map $g{\left(m, n\right)}$ is real-valued, the condition $c{\left(k_x, k_y\right)} = c^*{\left(-k_x, -k_y\right)}$ applies. 

We observe that the Fourier-space profiles are mostly concentrated around small $k_x$ and $k_y$, corresponding to the center region of the profiles in Fig. \ref{fig:mnist}(a). 
This is due to the slowly varying spatial features in the images. 
We can therefore expect that most of the information is carried by the small-$k$ Fourier components, and with the goal of decreasing the input size, we can restrict the data to $N$ coefficients with the smallest $k = \sqrt{k_x^2 + k_y^2}$. 
An additional advantage of this preprocessing step is that it reduces the computational resources required to perform the training process because the neural network dimension does not need to accommodate all $28^2 = 784$ pixel values as inputs.

Fourier preprocessing is particularly relevant for ONNs for two reasons. 
First, the Fourier transform has a straightforward implementation in the optical domain using techniques from Fourier optics involving standard components such as lens and spatial filters \cite{goodman_introduction_2005}. 
Second, this approach allows us to take advantage of the fact that ONNs are \textit{complex}-valued functions. 
That is to say, the $N$ complex-valued coefficients $c{\left(k_x, k_y\right)}$ can be handled by an $N$-dimensional ONN, whereas to handle the same input using a real-valued neural network requires a twice larger dimension. 
The ONN architecture used in our demonstration is shown schematically in Fig. \ref{fig:mnist}(a). 
The $N$ Fourier coefficients closest to $k_x = k_y = 0$ are fed into an optical neural network consisting of $L$ layers, after which a drop-mask reduces the final output to 10 components.  
The intensity of the 10 outputs are recorded and normalized by their sum, which creates a probability distribution that may be compared with the one-hot encoding of the digits from 0 to 9. 
The loss function is defined as the cross-entropy between the normalized output intensities and the correct one-hot vector.

During each training epoch, a subset of 60,000 images from the dataset were fed through the network in batches of 500. 
The remaining 10,000 image-label pairs were used to form a test dataset. 
For a two-layer network with $N = 16$ Fourier components, Fig. \ref{fig:mnist}(b) compares the classification accuracy over the training dataset (solid lines) and testing dataset (dashed lines) while Fig. \ref{fig:mnist}(b) compares the cross entropy loss during optimization. 
The blue curves correspond to an ONN with no activation function (e.g. a linear optical classifier) and the orange curves correspond to an ONN with the electro-optic activation function configured with $g_\phi = 0.05\pi$, $\phi_b = 1.00\pi$, and $\alpha=0.1$. 
The gain setting in particular was selected heuristically. 
We observe that the nonlinear activation function results in a significant improvement to the ONN performance during and after training. 
The final validation accuracy for the ONN with the activation function is $93\%$, which amounts to an 8\% difference as compared to the linear ONN which achieved an accuracy of $85\%$.

The confusion matrix computed over the testing dataset is shown in Fig. \ref{fig:mnist}(d).
We note that the prediction accuracy of $93\%$ is high considering that only $N = 16$ complex Fourier components were used, and the network is parameterized by only $2\times N^2 \times L = 1024$ free parameters. 
Moreover, this prediction accuracy is comparable with the $92.6\%$ accuracy achieved in a fully-connected linear classifier with 4010 free parameters taking \textit{all} of the $28^2 = 784$ real-space pixel values as inputs \cite{lecun_gradient-based_1998}. 
Finally, in Table \ref{tab:mnist_summary} we show that the accuracy can be further improved by including a third layer in the ONN and by making the activation function gain a trainable parameter. 
This brings the testing accuracy to $94\%$.
Based on the parameters from Table \ref{tab:parameters} and the scaling from Table \ref{tab:fom}, the 3 layer handwritten digit classification system would consume 4.8 W while performing $7.7 \times 10^{12}$ MAC/sec. Its prediction latency would be 1.5 ns.

\begin{table*}[t]

\caption{Accuracy on the MNIST testing dataset after optimization}

\setlength{\tabcolsep}{5pt}
\renewcommand{\arraystretch}{1.3}
\centering
\begin{tabular}{|c|c|cc|}
\hline
\textbf{\# Layers} & \textbf{Without activation} & \multicolumn{2}{c|}{\textbf{With activation}} \\
 &  & \textbf{Untrained} & \textbf{Trained}$^*$ \\ 
\hline
1 & 85.00\% & 89.80\% & 89.38\% \\
2 & 85.83\% & 92.98\% & 92.60\% \\
3 & 85.16\% & 92.62\% & 93.89\% \\
\hline
\end{tabular}
\\
\vspace{3pt}
\footnotesize{$^*$The phase gain, $g_\phi$, of each layer was optimized during training}

\label{tab:mnist_summary}
\end{table*}

\section{Conclusion}

In conclusion, we have introduced an architecture for synthesizing optical-to-optical nonlinearities and demonstrated its use as a nonlinear activation function in a feed forward ONN. 
Using numerical simulations, we have shown that such activation functions enable an ONN to be successfully applied to two machine learning benchmark problems: (1) learning a multi-input XOR logic function, and (2) classifying handwritten numbers from the MNIST dataset. 
Rather than using all-optical nonlinearities, our activation architecture uses intermediate signal pathways in the electrical domain which are accessed via photodetectors and phase modulators. 
Specifically, a small portion of the optical input power is tapped out which undergoes analog processing before modulating the remaining portion of the same optical signal. 
Whereas all-optical nonlinearities have largely fixed responses, a benefit of the electro-optic approach demonstrated here is that signal amplification in the electronic domain can overcome the need for high optical signal powers to achieve a significantly lower activation threshold. 
For example, we show that a phase modulator $V_\pi$ of 10 V and an optical-to-electrical conversion gain of 57 dB$\Omega$, both of which are experimentally feasible, result in an optical activation threshold of 0.1 mW.
We note that this nonlinearity is compatible with the \textit{in situ} training protocol proposed in Ref. \citenum{hughes_training_2018}, which is applicable to arbitrary activation functions.

Our activation function architecture can utilize the same integrated photodetector and modulator technologies as the input and output layers of a fully-integrated ONN. 
This means that an ONN using this activation suffers no reduction in processing speed, despite using analog electrical components. 
The only trade off made by our design is an increase in latency due to the electro-optic conversion process. 
However, we find that an ONN with dimension $N = 100$ has a total prediction latency of 2.4 ns/layer, with approximately equal contributions from the propagation of optical pulses through the interferometer mesh and from the electro-optic activation function. 
Conservatively, we estimate the energy consumption of an ONN with this activation function to be 100 fJ/MAC, but this figure of merit could potentially be reduced by orders of magnitude using highly efficient modulators and amplifier-free optoelectronics \cite{nozaki_femtofarad_2019}.

Finally, we emphasize that in our activation function, the majority of the signal power remains in the optical domain. 
There is no need to have a new optical source at each nonlinear layer of the network, as is required in previously demonstrated electro-optic neuromorphic hardware \cite{tait_neuromorphic_2017, peng_neuromorphic_2018, tait_silicon_2019} and reservoir computing architectures \cite{larger_photonic_2012, duport_fully_2016}. 
Additionally, each activation function in our proposed scheme is a standalone analog circuit and therefore can be applied in parallel. 
While we have focused here on the application of our architecture as an activation function in a feedforward ONN, the synthesis of low-threshold optical nonlinearlities using this circuit could be of broader interest for optical computing as well as microwave photonic signal processing applications.

\section*{Acknowledgments}
This work was supported by a US Air Force Office of Scientific Research (AFOSR) MURI project (Grant N\textsuperscript{\underline{o}} FA9550-17-1-0002). I.A.D.W. acknowledges helpful discussions with Avik Dutt.

\end{document}